\documentclass[12pt]{iopart}

\usepackage{graphicx}
\usepackage{iopams}
\begin{document}

\title[Modulational instability in photorefractives]{Modulational instability of optical beams in  photorefractive media in the presence of two wave and four wave mixing effects}

\author{C P Jisha,$^{1}$ V C Kuriakose,$^1$ K Porsezian $^2$ and B Kalithasan$^2$}

\address{$^1$ Department of Physics, Cochin University of Science and
Technology, Kochi-682 022}
\address{ $^2$ Department of Physics,
Pondicherry University, Pondicherry - 605014}

\ead{cpjisha@yahoo.co.in}

\begin{abstract}
Modulational instability in a photorefractive medium is studied in the presence of two wave mixing. We then propose and derive a model for forward four wave mixing in the photorefractive medium and investigate the modulational instability induced by four wave mixing effects. By using the standard linear stability analysis the instability gain is obtained. In both the cases,  the geometry is such that the effect of self-phase-modulation self-focusing is suppressed and only the holographic focusing nonlinearity is acting.

\end{abstract}

\maketitle

\section{Introduction}

A soliton is a localized wave that propagates without any change through
a nonlinear medium. Such a localized wave forms when the dispersion
or diffraction associated with the finite size of the wave is
balanced by the nonlinear change of the properties of the medium
induced by the wave itself \cite{mseg}. Solitons are ubiquitous in nature and
have been identified in various physical systems like fluids, plasmas,
solids, matter waves and classical field theory. Such self-trapped beams are the building blocks in future ultra-fast all-optical devices. They can be used to create reconfigurable optical circuits that guide other light
signals. Circuits with complex functionality and
all-optical switching or processing can then be achieved through the
evolution and interaction of one or more solitons \cite{sny}.
 It is possible to compress circuitry into a compact space with
many circuits sharing the same physical location. Furthermore,
certain photosensitive materials offer the potential for erasing one
light written device and replacing it by another. Hence, we have the
building blocks for dense reconfigurable virtual circuitry. Due to
the development of materials with stronger nonlinearities the
optical power needed to create such virtual circuits has been
reduced to the milliwatt and even microwatt level, bringing the
concept nearer to practical implementation.
 In a recent experimental study \cite{msh}, optical spatial screening
solitons have been observed for the first time in strontium barium
niobate (SBN). Soliton states of this sort are known to occur when
the process of diffraction is exactly balanced by light-induced
photorefractive (PR) waveguiding \cite{cmd}.

Modulational instability (MI) is a  process in
which tiny phase and amplitude perturbations that are
always present in a wide input beam grow exponentially
during propagation under the interplay between diffraction
(in spatial domain) or dispersion (in temporal domain) and
nonlinearity. Instabilities and chaos can occur in many types of nonlinear
physical systems. Optical instabilities can be classified as
temporal and spatial instabilities depending on whether the
electromagnetic wave is modulated temporally or spatially after it
passes through the medium. Temporal instability has been studied by
various authors \cite{hhe,gpa,ganapathy}, and the first experimental observation of MI
in a dielectric material was in 1986 \cite{ktai}. The temporal MI
occurs as an interplay between self-phase modulation and group
velocity dispersion. In optical fibers, MI occurs in the negative dispersion region and is responsible for the breakup into solitons. In the spatial domain, diffraction plays the
role of dispersion. When both diffraction and dispersion are present
simultaneously, it results in spatio-temproral MI \cite{lwl}.
Recently, Wen et al. investigated MI in negative refractive index
materials \cite{wen}. In a rather loose context, MI can be considered as a
precursor of self-trapped beam formation. In the spatial domain, MI
manifests itself as filamentation of a broad optical beam through
the spontaneous growth of spatial-frequency sidebands. The MI is a
destabilization mechanism for plane waves. It leads to
delocalization in momentum space and, in turn, to localization in
position space and the formation of self-trapped structures. During
MI, small amplitude and phase perturbations tend to grow
exponentially as a result of the combined effects of nonlinearity
and diffraction. This results in the disintegration of a large-diameter optical beam during propagation. Such instabilities have been widely studied in photorefractive media \cite{nan,cast,ms}.  Assanto et al. \cite{assanto} investigated  transverse MI  in undoped nematic liquid crystals, a highly nonlocal material system
encompassing a reorientational nonlinear response. They observed the one-dimensional development of transverse patterns which eventually lead to beam breakup, filamentation, and spatial soliton formation, using both spatially coherent and partially incoherent excitations. The MI phenomena has been previously observed in
various media like Kerr media \cite{she}, electrical circuits
\cite{bil}, plasmas \cite{has}, parametric band gap systems
\cite{hhe}, quasi-phase-matching gratings \cite{joel}, discrete dissipative systems \cite{ali} and in PR polymers \cite{cpprp}. The transverse instability of counterpropagating waves in PR media is studied by Saffman et al \cite{msa}. From the above investigations, it is clear that the study of MI in a medium is both of fundamental as well as of technological importance.

In this work, we investigate the modulational instability
induced in a photorefractive  medium  by two wave mixing  and the forward
four wave mixing  occurring in the PR material. The photorefractive medium is a well suited medium for the practical implementation of virtual circuits
as self-trapping of beams can be observed in this medium at very low
laser powers. Four wave mixing in the phase conjugate geometry is widely used in photorefractive materials for various applications. It can be used to implement several different computing functions \cite{pyehp}, optical
interconnects \cite{pyeh2}, matrix addition \cite{py}, and optical
correlator \cite{chal}. Nonlinear solutions for photorefractive
vectorial two-beam coupling and for forward phase conjugation in
photorefractive crystals have been found in \cite{bfish}. Recently Jia et al. \cite{jia} experimentally demonstrated degenerate forward
four wave mixing effects in a self-defocusing PR medium, in both one and two transverse dimensions. They observed the nonlinear evolution of new modes as a function of propagation distance, in both the near-field and far-field (Fourier space) regions.

The motivation behind studying MI in such a geometry is due to the recent proposal  by Cohen et al. \cite{oren} of a  new kind of spatial solitons, known as the holographic (HL) solitons. They are formed when the broadening tendency of diffraction is balanced by phase modulation that is due to Bragg diffraction from the induced grating. Holographic solitons are solely supported by cross-phase modulation arising from the induced grating, not involving self-phase modulation at all. In 2006 \cite{orenc}, they showed that the nonlinearity in periodically poled photovoltaic photorefractives can be solely of the cross phase modulation type. The effects of self-phase modulation and asymmetric energy exchange, which exist in homogeneously poled photovoltaic photorefractives, can be considerably suppressed by the periodic poling.
They demonstrated numerically that periodically poled photovoltaic photorefractives can support Thirring-type (solitons which exist only by virtue of cross phase modulation) ("holographic") solitons.  HL solitons in PR dissipative medium was studied by Liu \cite{liu}.  Existence of HL solitons in a grating mediated waveguide was studied by Freedman et al. \cite{freedman}. Salgueiro et al. \cite{salg} studied the composite spatial solitons supported by mutual beam focusing in a Kerr-like nonlinear medium in the absence of the self-action effects. They
predicted the existence of continuous families of single and
two-hump composite solitons, and studied their stability
and interaction. In 2007, two-dimensional holographic photovoltaic bright spatial solitons were observed in a
Cu:K$_{0.25}$Na$_{0.75}$Sr$_{1.5}$Ba$_{0.54}$Nb$_5$O$_{15}$ crystal in which two coherent laser beams, a signal beam, as well as
a strong and uniform pump beam at 532 nm are coupled to each other via two-wave mixing \cite{jins}.

This paper is organized as follows. The basic propagation equation for the two wave mixing geometry is presented in Section \ref{sec1} and the modulational instability in this geometry is studied. In section \ref{sec2} the governing equations for the forward four wave mixing geometry is presented. The system is studied without using the undepelted pump beam approximation. The standard linear stability analysis for the coupled equations is carried out and the gain spectrum is obtained.  Section \ref{con} concludes the paper.

\section{Two wave mixing geometry}\label{sec1}
In this section, we present the derivation of the model equation for the two wave mixing geometry \cite{pyeh, honda} and study the MI in it. Consider the
interaction of two laser beams inside a photorefractive medium. Stationary
interference pattern is formed, if the two beams are of the same
frequency.

Let the electric field of the two beams be written as
\begin{equation}
E_j = A_j\exp[i(\omega t - \textbf{k}_j.\textbf{r}
)], \textrm{for} ~ j=1,2.
\end{equation}
Here, $A_1 ~\textrm{and} A_2$ are the amplitudes, $\omega$ is the
angular frequency, and $\textbf{k}_1$ and $\textbf{k}_2$ are the wave
vectors.

The medium is assumed to be isotropic and both beams are polarized
perpendicular to the plane of incidence. The total intensity of the beams
\begin{equation}
I = |E|^2 = |E_1 + E_2|^2,
\end{equation}
 can be expressed as
\begin{equation}\label{int}
I = |A_1|^2 + |A_2|^2 + A_1A_2^*\exp[i\textbf{K}.\textbf{r}] +
A_2A_1^*\exp[-i\textbf{K}.\textbf{r}],
\end{equation} where
$$\textbf{K} = \textbf{k}_2 - \textbf{k}_1.$$
The magnitude of the vector $\textbf{K}$ is $2\pi/\Lambda$ where
$\Lambda$ is the period of the fringe pattern.

Equation \ref{int} represents a spatial variation of optical energy in the
 photorefractive medium. Such an intensity pattern will generate and redistribute
 charge carriers. As a result, a space charge field is created in
 the medium. This field induces an index volume grating via the
 Pockels effect. In general, the index grating will have a spatial
 phase shift relative to the interference pattern.

 The index of refraction including the fundamental component of the
 intensity-induced gratings can be written as
 \begin{equation}\label{refindext}
 n = n_0 + \left[\frac{n_1A_1^*A_2}{2I_0}\exp[i\phi]\exp[-i\textbf{K}.\textbf{r}] +
 \textrm{cc}\right],
 \end{equation}
 where $n_0$ is the index of refraction when no light is present,
 $\phi$ is real and $n_1$ is a real and positive number.
The wave equation reduces to the Helmholtz equation for highly
monochromatic waves like laser beams which can be viewed as a
superposition of many monochromatic plane waves with almost
identical wave vectors
 \begin{equation}\label{plnw}
 \nabla ^2 E + \omega ^2 n^2 E/c^2 = 0,
 \end{equation}
where $E = E_1 + E_2$.

Let $$E_j = A_j(x,y,z)\exp[i\omega t - i \beta _jz],$$

where $A(x,y,z)$ is the complex amplitude that depends on position,
$\beta_1$ and $\beta_2$ are the $z$ component of the wave vectors
$\textbf{k}_1$ and $\textbf{k}_2$ inside the medium respectively and
$z$ is measured along the central direction of propagation. We solve for the steady state so that $A_j$ is taken to be time independent, so that equation (\ref{plnw}) becomes
\begin{equation}\label{wave}
\Big(\frac{\partial^2}{\partial x^2} + \frac{\partial^2}{\partial y^2} +
\frac{\partial^2}{\partial z^2}  + \omega ^2 n^2 /c^2\Big)E = 0.
\end{equation}
Substituting equation (\ref{refindext}) in equation (\ref{wave}) and solving by neglecting the
second order term $n_1^2$ and using the fact that for highly
directional monochromatic waves with $a >> \lambda$, $\partial^2
A/\partial z^2 $
 can be neglected, we get
 \begin{equation}
 2i\beta _1 \frac{\partial A_1}{\partial z} = \nabla _\bot ^2 A_1 +\frac{\omega ^2 n_0 n_1}{c^2
 I_0}e^{-i\phi}A_2^* A_2 A_1,
 \end{equation}
and
 \begin{equation}
 2i\beta _2 \frac{\partial A_2}{\partial z} = \nabla _\bot ^2 A_2 + \frac{\omega ^2 n_0 n_1}{c^2
 I_0}e^{-i\phi}A_1^* A_1 A_2.
 \end{equation}

For the case when the two laser beams enter the medium from
the same side at $z=0$
$$\beta _1 = \beta _2 = k\cos\theta =
\frac{2\pi}{\lambda}n_0\cos\theta.$$

 where, $2\theta$ is the angle between the beams inside the medium. Simplifying  we obtain
\numparts
 \begin{eqnarray}\label{twmod}
 i\frac{\partial A_1}{\partial z} = \frac{1}{2k \cos\theta}\nabla _\bot ^2 A_1
 +\frac{\Gamma}{2 I_0}A_1| A_2|^2,\\
 i\frac{\partial A_2}{\partial z} = \frac{1}{2k \cos\theta}\nabla _\bot ^2 A_2 + \frac{\Gamma}{2 I_0}A_2| A_1|^2,
 \end{eqnarray}
 \endnumparts
where \begin{equation}\Gamma = \frac{2\pi n_1}{\lambda \cos\theta}e^{-i\phi}\end{equation} is
the complex coupling coefficient.  A similar model was used to study the existence of Holographic solitons \cite{oren}.
A prerequisite for obtaining Holographic focusing is that the induced grating be in phase with the intensity grating. If the induced grating is shifted by $\pi$, then the grating leads to holographic defocusing. If the grating is $\pm \pi/2$ phase shifted with respect to the intensity grating, then the interaction will yield an asymmetric energy exchange between the two beams because of the two beam coupling property of the photorefractive material. Therefore, $\phi = 0$ for the existence of bright solitons. This gives the coupling constant as
\begin{equation}
\Gamma = \frac{2\pi n_1}{\lambda \cos\theta}.
\end{equation}

The next step is to study the propagation of a broad optical beam through a PR medium. We study MI of a one dimensional broad optical beam. Hence the $y$ dependent term in the transverse Laplacian in equation (\ref{twmod}) can be neglected. For a broad optical beam, the diffraction term in equation (\ref{twmod} ) can be set to zero giving the steady state solutions as
\begin{equation}\label{stdt}
A_j = \sqrt P \exp\Big[-i\frac{\Gamma P}{2 I_0}z\Big],
\end{equation}

for $j = 1,2$.

To study MI, we consider small perturbations of the steady state solutions :
\begin{equation}\label{pstdt1}
A_j = (\sqrt P + a_j) \exp\Big[-i\frac{\Gamma P}{2 I_0}z\Big].
\end{equation}

Substituting  in (\ref{twmod}) and neglecting the quadratic and higher order terms in $a_j$, the perturbations $a_1$ and $a_2$ are found to satisfy the following linearized set of two coupled equations :
\begin{equation}\label{per1}
i \frac{\partial a_1}{\partial z} = \frac{1}{2k \cos\theta} \frac{\partial ^2 a_1}{\partial x^2} + \frac{\Gamma P}{2 I_0}(a_2 + a_2^*),
\end{equation}

\begin{equation}\label{per2}
i \frac{\partial a_2}{\partial z} = \frac{1}{2k \cos\theta} \frac{\partial ^2 a_2}{\partial x^2} + \frac{\Gamma P}{2 I_0}(a_1 + a_1^*).
\end{equation}

It is important to note that the evolution of the perturbations depend solely on the cross phase modulation. To solve equations (\ref{per1}) and (\ref{per2}), we assume that the perturbations be composed of two side bands :
\begin{equation}\label{sdbt}
 a_j(x,z) = U_j(z)\exp [i\kappa x] + V_j(z)\exp [-i\kappa x].
 \end{equation}

The substitution of equation (\ref{sdbt}) in  (\ref{per1}) and (\ref{per2}) results in a set of four homogeneous equations in $U_1$, $U_2$, $V_1$ and $V_2$ as
\begin{eqnarray}
&&\frac{\partial U_1}{\partial z} = i \frac{\kappa ^2}{2k \cos \theta}U_1  - i\frac{\Gamma P}{2I_0}(U_2 + V_2^*),\\ &&\frac{\partial V_1^*}{\partial z} = -i \frac{\kappa ^2}{2k \cos \theta}V_1^*  + i\frac{\Gamma P}{2I_0}(U_2 + V_2^*),\\
&&\frac{\partial U_2}{\partial z} = i \frac{\kappa ^2}{2k \cos \theta}U_2  - i\frac{\Gamma P}{2I_0}(U_1 + V_1^*),\\
&&\frac{\partial V_2^*}{\partial z} = -i \frac{\kappa ^2}{2k \cos \theta}U_1  + i\frac{\Gamma P}{2I_0}(U_1 + V_1^*).
\end{eqnarray}

\begin{figure}
 \centering
\includegraphics{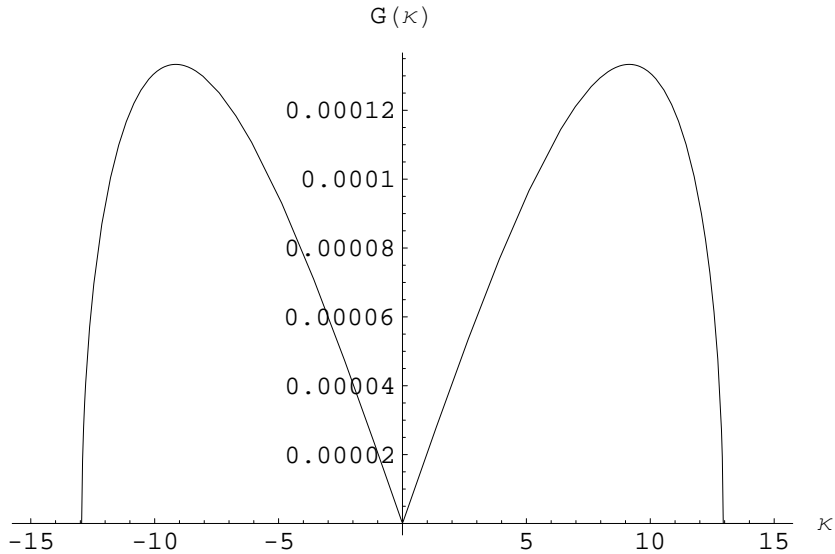}
\caption{Growth rate as a function of spatial frequency for
the two wave mixing geometry. }\label{twm1}
\end{figure}

\begin{figure}
 \centering
\includegraphics{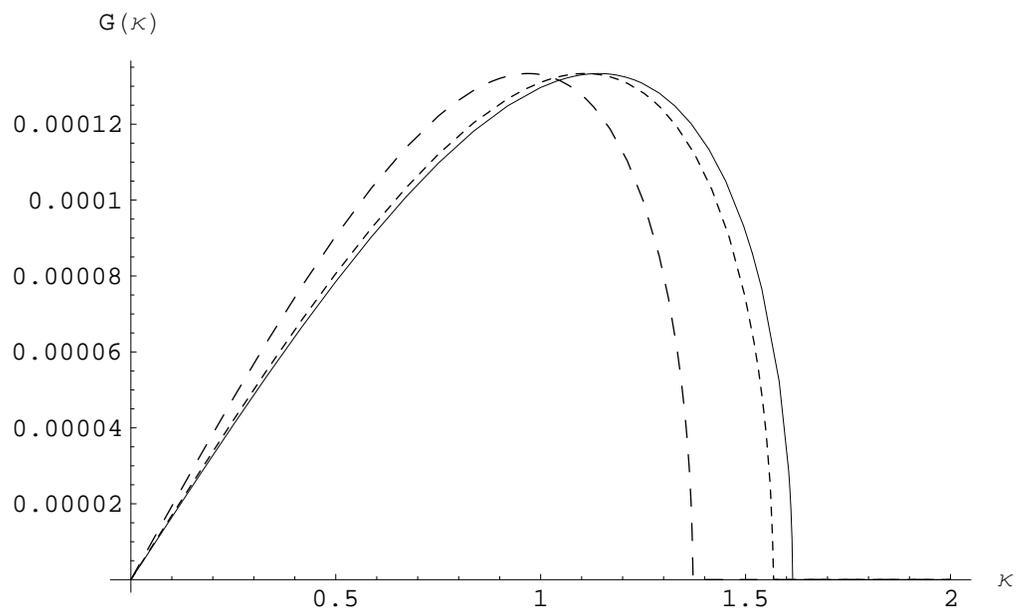}
\caption{Variation of gain with respect to the angle between the two beams. The lowermost curve is for $\theta = \pi/16$, the dotted curve is for $\pi/8$ and the topmost curve is for $\pi/4$.   }\label{miang}
\end{figure}

This set has a nontrivial solution only if the determinant of the coefficient matrix vanishes. The eigenvalues of the system are obtained as \begin{eqnarray}
\Lambda_{1\pm} = \pm \frac {1}{ 2A} \Big(\frac{- I_0 \kappa^4 - 2 A \Gamma P \kappa^2}{ I_0}\Big)^{1/2},\\
\Lambda_{2\pm} = \pm \frac {1}{ 2A} \Big(\frac{- I_0 \kappa^4 + 2 A \Gamma P \kappa^2}{ I_0}\Big)^{1/2},
\end{eqnarray}
where $A = k \cos \theta$. The plane wave solution is stable if perturbations at any wave number $\kappa$ do not grow with propagation.  MI gain will exist only when $Re[\Lambda] > 0$. This condition is satisfied only by the second set of eigenvalues.
The gain associated with the system is given by
\begin{equation}
\textbf{G} = |\Re(\Lambda_\pm)|.
\end{equation}
A typical plot of the gain spectrum is given in figure (\ref{twm1}).  We consider the case of BaTiO$_3$ crystal with a space charge field of $10^4$V/m, electro-optic coefficient $r_{42} = 1640 \times 10^{-12}$m/V and refractive index $n = 2.4$ which gives a coupling constant $\Gamma = 20$ cm$^{-1}$. Figure \ref{miang} gives the variation of gain with  the angle $\theta$. The gain increases with decrease in angle.

\section{Forward four wave mixing geometry}\label{sec2}
\begin{figure}
\centering
\includegraphics{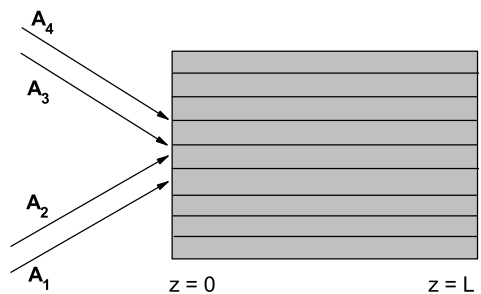}
\caption{Forward four wave mixing in photorefractive media in the transmission geometry. }\label{fwmgeo}
\end{figure}
In the two wave mixing case, two coherent beams interfere inside a
photorefractive medium and produce a volume index grating. In the
case of optical phase conjugation, using four-wave mixing, a third
beam is incident at the Bragg angle from the opposite side and a
fourth beam is generated. We now consider the situation in which the third beam is incident from the front at the Bragg angle and a diffracted beam is
generated. That is we consider the interaction of four beams in a PR
medium in the forward geometry (see figure (\ref{fwmgeo})). We assume that all the beams have
the same frequency $\omega$. We propose a model for the observation
of MI in a PR medium induced by four wave mixing. The method
proceeds by first writing the four coupled equations for the present
geometry. Of the four beams, let beams $A_2$ and $A_3$ be the pump beam, $A_1$ be the signal beam and $A_4$ be the generated beam. Beam $A_1$ is coherent with beam $A_2$ and beam $A_3$ is coherent with beam $A_4$. Thus the index grating consists of two contributions : $A_1^*A_2$ and $A_3^*A_4$. The index of refraction including the fundamental component of the intensity-induced gratings can thus be written as
 \begin{equation}\label{refindex}
 n = n_0 + \left[\frac{n_1(A_1^*A_2 + A_3^*A_4)}{2I_0}\exp[i\phi]\exp[-i\textbf{K}.\textbf{r}] +
 \textrm{cc}\right],
 \end{equation}
where $\textbf{k}_2 - \textbf{k}_1= \textbf{k}_4 - \textbf{k}_3 = \textbf{K}$.

 Following the above procedure for the two wave mixing case, we get the following four
coupled equations for the present geometry:
\numparts\label{cpfwm}
\begin{eqnarray}
 i\frac{\partial A_1}{\partial z} = \frac{1}{2k \cos \theta}\nabla _\bot ^2 A_1
 +\frac{\Gamma}{2 I_0}(A_1A_2^* + A_3A_4^*)A_2,\\
 i\frac{\partial A_2}{\partial z} = \frac{1}{2k \cos \theta}\nabla _\bot ^2 A_2 + \frac{\Gamma}{2
 I_0}(A_1^*A_2 + A_3^*A_4)A_1,\\
 i\frac{\partial A_3}{\partial z} = \frac{1}{2k \cos \theta}\nabla _\bot ^2 A_1
 +\frac{\Gamma}{2 I_0}(A_1A_2^* + A_3A_4^*)A_4,\\
 i\frac{\partial A_4}{\partial z} = \frac{1}{2k \cos \theta}\nabla _\bot ^2 A_2 + \frac{\Gamma}{2
 I_0}(A_1^*A_2 + A_3^*A_4)A_3,
 \end{eqnarray}
 \endnumparts where
$I_0 = I_1 + I_2 + I_3 + I_4$.

 The model permits plane wave solutions of the form $A_j(x,z) = \sqrt  P \exp [-i\frac{\Gamma P}{2 I_0}z]$. The next step is to carry out a Linear Stability Analysis of
 the plane wave solutions. For this the plane wave solution is perturbed as
 \begin{equation}
A_j = (\sqrt P + a_j(x,z)) \exp\Big[-i\frac{\Gamma P}{2 I_0}z\Big],
 \end{equation}
 where $a_j$ is a small complex perturbation. Inserting this into
 the  coupled equation (\ref{cpfwm})  and linearizing around the solution yields the equations for the perturbations:
 \begin{equation}
i\frac{\partial a_1}{\partial z} = -\frac{\Gamma}{2
I_0} a_1 +
\frac{1}{2k \cos \theta}\frac{\partial ^2 a_1}{\partial x^2} + \frac{\Gamma}{2
I_0}(2a_2 + a_2^* +a_3 + a_4),
 \end{equation}

\begin{equation}
i\frac{\partial a_2}{\partial z} = -\frac{\Gamma}{2
I_0} a_2 +
\frac{1}{2k \cos \theta}\frac{\partial ^2 a_2}{\partial x^2} + \frac{\Gamma}{2
I_0}(2a_1 + a_1^* + a_3^* + a_4),
 \end{equation}

 \begin{equation}
i\frac{\partial a_3}{\partial z} = -\frac{\Gamma}{2
I_0} a_3 +
\frac{1}{2k \cos \theta}\frac{\partial ^2 a_3}{\partial x^2} + \frac{\Gamma}{2
I_0}(a_1  + a_2^* + 2a_4 + a_4^*),
 \end{equation}

 \begin{equation}
i\frac{\partial a_4}{\partial z} = -\frac{\Gamma}{2
I_0} a_4 +
\frac{1}{2k \cos \theta}\frac{\partial ^2 a_4}{\partial x^2} + \frac{\Gamma}{2
I_0}(a_1^* + a_2 + 2a_3 + a_3^*).
 \end{equation}

 Now, we assume that the spatial perturbation $a(x,z)$ is
 composed of two side band plane waves, i.e
 \begin{equation}
 a_j(x,z) = U_j(z)\exp [i\kappa x] + V_j(z)\exp [-i\kappa x].
 \end{equation}

Substituting, we get eight  homogeneous equations as
 \numparts \label{ehe}
\begin{eqnarray}
&&\frac{\partial U_1}{\partial z} = - i (-\frac{\Gamma P}{2 I_0} - \frac{\kappa^2}{2k\cos \theta})U_1 - i\frac{\Gamma P}{2 I_0}(2U_2 + V_2^* + U_3 + V_4^*),\\ &&\frac{\partial U_1}{\partial z} = - i (-\frac{\Gamma P}{2 I_0} - \frac{\kappa^2}{2k\cos \theta})U_2 - i\frac{\Gamma P}{2 I_0}(2U_1 + V_1^* + V_3^* + U_4),\\
&&\frac{\partial U_3}{\partial z} = - i (-\frac{\Gamma P}{2 I_0} - \frac{\kappa^2}{2k\cos \theta})U_3 - i\frac{\Gamma P}{2 I_0}(U_1 + V_2^* + 2U_4 + V_4^*),\\
&&\frac{\partial U_4}{\partial z} = - i (-\frac{\Gamma P}{2 I_0} - \frac{\kappa^2}{2k\cos \theta})U_4 - i\frac{\Gamma P}{2 I_0}(V_1^* + U_2 + 2U_3 + V_3^*),\\&&\frac{\partial V_1^*}{\partial z} = - i (-\frac{\Gamma P}{2 I_0} - \frac{\kappa^2}{2k\cos \theta})V_1^* + i\frac{\Gamma P}{2 I_0}(2V_2^* + U_2 + V_3^* + U_4),\\&&\frac{\partial V_2^*}{\partial z} = - i (-\frac{\Gamma P}{2 I_0} - \frac{\kappa^2}{2k\cos \theta})V_2^* + i\frac{\Gamma P}{2 I_0}(2V_1^* + U_1 + U_3 + V_4^*),\\&&\frac{\partial V_3^*}{\partial z} = - i (-\frac{\Gamma P}{2 I_0} - \frac{\kappa^2}{2k\cos \theta})V_3^* - i\frac{\Gamma P}{2 I_0}(V_1^* + U_2 + 2V_4^* + U_4),\\&&\frac{\partial V_4^*}{\partial z} = - i (-\frac{\Gamma P}{2 I_0} - \frac{\kappa^2}{2k\cos \theta})V_4^* - i\frac{\Gamma P}{2 I_0}(U_1 + V_2^* + 2V_3^* + U_3).
\end{eqnarray}
\endnumparts

The obtained eight coupled equations  can be written in the compact
matrix form as, $\partial _z \underline{X} =  \mathbf{M}
\underline{X}$, where $\mathbf{M}$ is an 8x8 matrix with $\underline{X}$  = [$U_1$ $U_2$ $ U_3$ $U_4$ $V_1^*$ $V_2^*$ $V_3^*$ $ V_4^*$]$^T$.

This system has a nontrivial solution only if the determinant of
the matrix vanishes. The real part of the eigenvalues of the
stability matrix of equation (\ref{ehe}) gives the gain associated with
the system \cite{boyd}. Out of the eight roots of the system, only the root with maximum positive value contributes to the MI gain. Here only the eigenvalue given by

\numparts\label{line}
\begin{eqnarray}
\Lambda_\pm = \pm \frac {\sqrt{- \kappa^4 I_0 + 4 \kappa^2 \Gamma k \cos \theta P}}{2 \sqrt{I_0}k \cos\theta}\end{eqnarray}
\endnumparts

contributes to the MI of the system.

We note that the gain vanishes for all values of $\kappa$ greater than $\kappa_{max} = 4k\Gamma P/I_0$. Defining $\gamma = \Gamma P/I_0 $, we can rewrite the gain as

\begin{equation}
\frac{G}{\gamma} = \sqrt{4 \Big(\frac{\kappa}{\kappa_{max}}\Big)^2 \Big(1- \Big(\frac{\kappa}{\kappa_{max}}\Big)^2\Big)}.
\end{equation}

\begin{figure}
\centering
\includegraphics[width = 8cm]{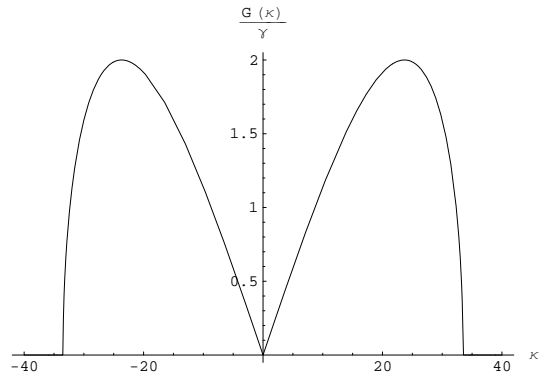}
\caption{A typical plot showing the gain spectrum of the system
in the forward four wave mixing process with respect to the spatial perturbation $\kappa$. }\label{fig}
\end{figure}

The variation of the gain coefficient in the forward four wave mixing process with respect to the spatial perturbation is plotted in figure (\ref{fig}). Such instabilities are useful for pattern formation. A transverse modulation instability of a single beam or counterpropagating beams is a general mechanism that leads to pattern formation in nonlinear optics  \cite{mschc,osand}. We expect that similar results will be obtained using the present geometry.

\section{Conclusion}\label{con}

We first studied MI occurring in a PR medium in the two wave mixing geometry and further modeled the forward four wave mixing occurring in the PR medium and studied MI in this geometry. In both the cases,  the geometry is such that only the holographic focusing nonlinearity is acting. MI does not rely on self-phase-modulation self-focusing but results  only by virtue of the  competition between induced periodic modulation of the refractive index and diffraction of the beam.  The MI gain spectrum is obtained for both two wave mixing and forward four wave mixing geometry. Such instabilities will be useful for pattern formation. Photorefractive materials are attractive for the studies of pattern formation as their slow time constant gives the possibility of observing the spatiotemporal dynamics of the system in real time. This also reduces the demands on experimental equipment where speed is often a crucial parameter. Another advantage of photorefractive pattern formation is that patterns can be observed using optical powers of tens of mW. In contrast, pattern formation using other nonlinearities require optical powers in the order of 1W.

\ack
J.C.P and V.C.K wish to
acknowledge DST (Research Grant) for financial
support. J.C.P also acknowledges CSIR for the award of senior
research fellowship. KP wishes to thank DST, DAE-BRNS, IFCPAR, CSIR and DST-Ramanna fellowships for financial help through
projects.


\end{document}